\documentclass[10pt]{article}
\title{{\bf Discovering Algorithms with Matrix Code}}

\author{M.H. van Emden \\
        Department of Computer Science \\
        University of Victoria, Canada \\
        Research Report DCS-345-IR
}

\date{}

\begin{document}
\maketitle
\begin{abstract}
In first-year programming courses it is often difficult to show
students how an algorithm can be discovered.  In this paper
we present a program format that supports the development from
specification to code in small and obvious steps;
that is, a \emph{discovery process}.
The format, called \emph{Matrix Code}, 
can be interpreted as a proof according to
the Floyd-Hoare program verification method.
The process consists of expressing the specification
of a function body as an initial \emph{code matrix},
and then growing the matrix by adding rows and columns
until the completed matrix is translated in a routine
fashion to compilable code.
As worked example we develop a Java program that generates
the table of the first $N$ prime numbers.

\end{abstract}

\newcommand{\emln}{$\;$\\} 
\newcommand{\lmnt}[1]{\parbox{1.0in}{
   {{\emln \tt #1 }} \emln}}
\newcommand{\lmntWdth}[2]{\parbox{#1in}{
    {{\emln \tt #2 }} \emln}}
\newcommand{\nc}[1]{\lmntWdth{0.6}{#1}} 
\newcommand{\emp}{\lmnt{$\;$}} 

\newcommand{\set}[2]{\{#1 : #2\}}
\newcommand{\bk}[3]
  {{\tt \{#1\}#2\{#3\}}}
\newcommand{\brr}[2]{{\tt \{#1\}#2}}
\newcommand{\ktt}[2]{#1\{#2\}}

\newcommand{\vc}[3]{${\mathbf V}_{\mbox{{\tt #2}}}(#1;#3)$}
\newcommand{\trpl}[3]{{\tt \{#1\}#2\{#3\}}}
\newcommand{\mpr}[2]{{\tt #1:#2}} 
\newcommand{\mst}[2]{{\tt (#1,#2)}} 

\newcommand{\mbf}[1]{\noindent {\bf #1}} 

\section{Introduction}

Structured programming was a great step forward
from the preceding stage of chaotic programming.
The next step, beyond structured programming,
is \emph{verification-driven} programming,
where proof of correctness and code are developed in parallel.
Matrix Code is important for the practicing programmer
because it makes verification-driven programming possible.
But it also helps to solve the problem
faced by an instructor in a first-year programming class:
most of the class understands why a given program works,
but how to help the student,
faced with a blank screen,
\emph{to get started} with the program of the next
assignment?
This is where Matrix Code helps:
there is always something (small) to do,
and when there is nothing more to do,
the code matrix is ready for routine translation to Java
with confidence that the resulting code
has the desired behaviour.

E.W. Dijkstra addressed the same problem when he started
teaching in the 1960's.
His remedy was a detailed step-by-reasoning
and construction resulting in
an Algol-60 program for filling an array with the first
thousand prime numbers.
In this paper we address the same task
to ease comparison with Dijkstra's report \cite{djkddh72}.

\section{Hoare's verification method}
\label{sec:floyd}

As an introduction to the verification method
due to R.W. Floyd and C.A.R. Hoare
we verify a Java version of the prime-number
generating program
developed by Dijkstra in \cite{djkddh72}.
See Figure~\ref{fig:floyd}.

\begin{figure}[htbp]
\begin{center}
\hrule \vspace{0.1in}
\begin{verbatim}
public static void primes(int[] p, int N) {
  // S
  int j,k,n;
  p[0] = 2; p[1] = 3; k = 2;
  // A
  while (k<N) {
    j = p[k-1]+2; n = 0;
    // B
    while (p[n]*p[n] <= j) {
      // C
      if (j%p[n+1] != 0) n++;
      else {j += 2; n = 0;}
    }
    p[k++] = j;
  }
  // H
}
\end{verbatim}
\end{center}
\caption{\label{fig:floyd}
An example of a Java function for filling
{\tt p[0..N-1]} with the first {\tt N} primes.
At the points indicated by the comments S, A, B, C, H
we need assertions to allow verification by Hoare's method.
}
\vspace{0.1in}
\hrule
\end{figure}

The essence of imperative code is that computation
progresses through the code along a well-defined
set of \emph{code locations}.
In Figure~\ref{fig:floyd} some of these locations
are indicated by the comments
\verb"S", \verb"A", \verb"B", \verb"C", and \verb"H".

We think of a computation as a sequence of
\emph{computation states}
each of which consists of a \emph{control state}
(a code location)
and a \emph{data state}
(a vector of values of the variables).

According to the Floyd-Hoare method,
assertions are attached to selected code locations.
The assertions assert that certain relations
between program variables hold at the code locations concerned.
When such an assertion occurs in a loop,
it is the familiar \emph{invariant} of that loop.
In Figure~\ref{fig:floyd} we have indicated by the comments
where these assertions have to be placed. 
Figure~\ref{fig:verification} contains the corresponding
assertions and the required Hoare triples
(see following explanation).

\begin{figure}[htbp]
\hrule
\vspace{0.1in}
\begin{verbatim}
Assertions:
S: p[0..N-1] exists and N>1
H: p[0..N-1] are the first N primes
A: S && p[0..k-1] are the first k primes && k <= N
B: A && k<N && relB(p, k, n, j)
C: B && p[n]*p[n] <= j

relB(p,k,n,j)} means that there is no prime
between p[k-1] and j, and that j is not divided
by any prime in p[0..n], and that n<k.

Hoare triples:

{S} p[0]=2; p[1]=3; k=2; {A}
{A && k >= N}  {H}
{A && k < N} j=p[k-1]+2; n=0; {B}
{B && p[n]*p[n] <= j} {C} 
{B && p[n]*p[n] > j} p[k++] = j {A}
{C && j%p[n+1] != 0} n++ {B}
{C && j%p[n+1] == 0} j += 2; n = 0 {B}
\end{verbatim}
\caption{\label{fig:verification}
Assertions and Hoare triples for Figure~\ref{fig:floyd}.
The meaning of a Hoare triple {\tt \{A0\} CODE \{A1\}}
is that if assertion {\tt A0} is true
and if {\tt CODE} is executed
with termination, then assertion {\tt A1} is true.
}
\vspace{0.1in}
\hrule
\end{figure}

The verification of the function as a whole
relies on the verification of a number of
implications defined in terms of assertions
and program elements such as tests and statements.
Consider Figure~\ref{fig:floyd}:
because there is an execution path
from \verb+A+ to \verb+B+, one has to show
the truth of\\
\verb"        {A && k<N} j=p[k-1]+2; j=0; {B}"\\
It has as meaning:
if \verb"A && k<N" (the \emph{precondition}) is true
and if\\
\verb"        j=p[k-1]+2; j=0;"\\
is executed, then \verb"B" (the \emph{postcondition})
is true.
Because of the three elements: precondition,
postcondition, and the item in between,
this is called a \emph{Hoare triple}.

There are many other implementations of the function
in Figure~\ref{fig:floyd} that are verified by the
same set of triples as in Figure~\ref{fig:verification}.
It would be tempting to say that,
once we have a sufficient set of Hoare triples,
we can forget the program in Figure~\ref{fig:floyd}:
all information about it is in the Hoare triples
of Figure~\ref{fig:verification}.
This may seem so because, for example, in\\
\verb"        {A && k < N} j=p[k-1]+2; n=0; {B}"\\
{\tt A} stands for the assertion defined earlier in that figure.
What is missing is the fact that assertion \verb"A"
is tied to code location {\tt A}.
But the idea of regarding the set of triples as the
essence is a fruitful one.
It leads one to ask:
what is a format for the information in
Figure~\ref{fig:verification} plus
the fact that the code locations are tied
to the assertions of the same name?
For the answer to this question we propose
Matrix Code.

\section{Matrix Code}
A code matrix can be thought of as
a graph with code locations as nodes
and directed arcs that are labeled with code.
A natural notation for such a graph is a matrix
with columns and rows labeled by nodes
and the arc labels as matrix entries.
When we use this notation for Figure~\ref{fig:verification},
then we get the matrix in 
Figure~\ref{fig:primes3}.
This is a \emph{code matrix}.

Although a code matrix arose from a collection of logical
statements, it can also be interpreted as specifying
a set of computations of an abstract machine.
In the explanation below we use a picturesque terminology
that is worth trying on an audience of novices in programming.

A code matrix is executed by a \emph{turtle}
moving over it.
The turtle contains a data state and a knowledge state.
The data state is a vector of the values of the variables
accessible from the code under construction.
The knowledge state is an assertion concerning
the values of these variables.

The turtle has an innate truthfulness
that prevents it from knowing a lie.
In other words,
its knowledge state is always
an assertion that is true of its data state.
The turtle is a logical animal
in the sense that it is endowed with an innate drive
that makes it draw a conclusion from the assertion
that is its knowledge state
and from it data state.
And the turtle only has available for its
conclusions the Hoare triples of the code matrix
and its own data and knowledge states.

The knowledge state specifies a row or column of the code matrix.
In Figure~\ref{fig:primes3} the knowledge state
can have as values \verb"S", \verb"A", \verb"B",
\verb"C", and \verb"H".
The data state is a vector of the values of the variables.
In Figure~\ref{fig:primes3} the data state
has as components the content of array \verb"p"
and the values of the variables \verb"k", \verb"j", and \verb"n".
We call the entries of the code matrix \emph{gates}.

Execution of a code matrix consists of 
the turtle performing a sequence of cycles.
The turtle's data and knowledge states are
updated as a consequence of executing the cycle.
At the beginning of the cycle the turtle
enters from the top of the matrix
through the column indicated by
the current knowledge state
until it encounters a gate.
The data state passes the gate or fails to do so.
In the latter case execution terminates with failure.
If the data state passes through the gate,
then the turtle exits the matrix to the right
through the row in which that gate occurs.
The new state has the label of that row
as knowledge state and as data state the one determined by
having had to pass through the gate.
This completes the cycle.

Initially the turtle has knowledge state {\tt S}.
When the knowledge state changes to {\tt H},
execution halts with success.

What determines whether the turtle can pass
through a gate and how does its state change when it does?
If a gate is a boolean expression that evaluates to {\tt false}
in the data state,
then the turtle fails to pass.
If it evaluates to \verb"true",
then the data state passes and remains unchanged.
If a gate is an assignment statement,
then the turtle passes if execution
of the statement is defined and terminates.
When it passes, then the data state is changed as
defined by the semantics of the statement.
In Figure~\ref{fig:primes3} we see
that gates may be composed by means of a semicolon.
In general, if $g$ and $h$ are gates, then
$g;h$ is also a gate.
The data state $s$ passes through
gate $g;h$ yielding state $t$
if it passes through gate $g$ giving $s'$
and if $s'$ passes through gate $h$ giving $t$.

The concept of gate is especially useful
because of the possibility that a gate consisting of
a boolean expression can be composed with an assignment.
Such a gate may block the data state because
the boolean expression evaluates to \verb"false".
When the data state is not thus blocked,
it will be transformed by the assignment statement.
Gates may be composed of boolean expressions
and statements in any order.
For ease of translation to a conventional language like Java,
we ensure that boolean expressions precede statements.

For example, suppose we start execution of Figure~\ref{fig:primes3}
with parameter {\tt N} equal to 1000
and with knowledge state equal to {\tt S}.
As there is only one triple in column {\tt S},
and as this triple occurs in row {\tt A},
the computation continues with column {\tt A}.
Only one of the gates in that column
allows the data state to pass,
so computation continues in the row of that gate,
namely {\tt B}.
Here follows an excerpt of the computation:

\begin{center}
\begin{verbatim}

N = 1000

knowledge | data state
state     | 
          | k     j      n    p
       -----------------------------------------
       S  |
       A  | 2                 {2,3,...}
       B  | 2     5      0    {2,3,...}
       C  | 2     5      0    {2,3,...}
       B  | 2     5      1    {2,3,...}
       A  | 3     5      1    {2,3,5,...}
      ... |        . . . 
       H  | 1000  7919   23   {2,3,5,...,7919}
\end{verbatim}
\end{center}

\section{Algorithm discovery from first principles}
\label{sec:ratio}

Let us use Matrix Code
to discover an algorithm for filling an array \verb"p[0..N-1]"
with the successive prime numbers
$
p_0 = 2,
p_1 = 3,
p_2 = 5,
\ldots,
p_{N-1}.
$

The specification of the desired function body \verb"G"
can be given as the Hoare triple \verb"{S}G{H}"
with {\tt S} and {\tt H} as in Figure~\ref{fig:primes0}.
This triple becomes part of the final code matrix;
see Figure~\ref{fig:primes0}.
As we don't have an immediate implementation
of gate \verb"G" in Java, we need to expand the code matrix.
One by one we add rows and columns
in such a way that the matrix is expanded
from the top right corner downward and to the left.

\begin{figure}[htbp]
\begin{tabular}{l|l||l}
\lmnt{} & \lmnt{S: p[0..n-1] exists \&\& n>1} & \\
\hline \hline
        & \lmnt{/*which G?*/}
        & \lmnt{H: p[0..n-1] contains the first n primes}  \\
\hline
        & & \lmnt{  \\ \\}                       \\

\end{tabular}
\caption{\label{fig:primes0}
A code matrix solving the problem,
if only we had an easy implementation for gate {\tt G}
such that $\bk{S}{G}{H}$.
We need at least one intermediate assertion;
see Figure~\ref{fig:primes1}.
}
\end{figure}

Assertion {\tt H} is too ambitious to achieve
with a simple gate when the data state is as
described by {\tt S}.
So we need at least one condition,
say, {\tt A}, that is intermediate between {\tt S}
and {\tt H} in the sense
that \bk{S}{G1}{A} and \bk{A}{G2}{H}
for simple {\tt G1} and {\tt G2}.
Less formally: if we can't fill all of a prime-number table
of size $N$ right away,
we can at least fill a small one,
say, of size $k \leq N$.
This suggests as intermediate assertion \verb"A":
the first {\tt k} primes in increasing order are in
{\tt p[0..k-1]} with {\tt 1 < k <= N}.
It is easy to reach
\verb"A"
from
\verb"S":
we put the first two primes in the table
and set \verb"k=2".

This allows us to exit through the new row for {\tt A},
setting the knowledge state to {\tt A}.
In the next step we enter through the column determined by
the knowledge state, hence column {\tt A}.
If the knowledge state is {\tt A}
(and if that assertion holds),
then passing the gate {\tt k >= N}
allows us to halt by exiting through row {\tt H}.
The new row and column update our code matrix to
the one in Figure~\ref{fig:primes1}.

\begin{figure}[htbp]
\begin{center}
\begin{tabular}{l|l|l||l}
\lmnt{} & \lmnt{A:} & \lmnt{S: p[0..N-1] exists \&\& N>1} & \\
\hline \hline
& \lmnt{k >= N} & & \lmnt{H: p[0..N-1] contains the first N primes}  \\
\hline
& & \lmnt{p[0] = 2; p[1] = 3; k = 2}
   & \lmnt{A: p[0..k-1] contains the first k primes \&\&
k <= N}  \\
\hline
        &  & & \lmnt{  \\ \\}                       \\
\end{tabular}
\caption{\label{fig:primes1}
Next step after Figure~\ref{fig:primes0}:
in column $A$ the case {\tt k < N} is missing.
This leads to a new row and column labeled {\tt B}
in Figure~\ref{fig:primes2}.
}
\end{center}
\end{figure}

However, when execution enters through column {\tt A}
in Figure~\ref{fig:primes1},
we may have that {\tt k < N},
so that we do not pass the gate {\tt k >= N}.
This points to the need to increase {\tt k},
hence to find the next prime after {\tt p[k-1]}.
Let {\tt j} be the candidate for this next prime.
That suggests including in assertion {\tt B:}
``{\tt A} is true and {\tt k<N} and {\tt j} is such that
there is no prime greater than {\tt p[k-1]}
and less than {\tt j} and {\tt j} is not divisible by any
of {\tt p[0..n]}'',
a statement that we abbreviate to {\tt relB(p,k,n,j)},
as in Figure~\ref{fig:verification}.

Column {\tt A} is now completed with a gate
allowing exit through the new row for {\tt B}.
When we enter through the new column for {\tt B}
we immediately know one of the gates
in column {\tt B} for the easy case
where the candidate {\tt j} for the next prime
actually turns out to be the next prime.
See Figure~\ref{fig:primes2} for
the resulting stage in the development of the code matrix.

\begin{figure}[htbp]
\begin{center}
\begin{tabular}{|l|l|l||l}
\lmnt{B:} & \lmnt{A:} & \lmnt{S: p[0..N-1] exists \&\& N>1} & \\
\hline \hline
& \lmnt{k >= N} & & \lmnt{H: p[0..N-1] contains the first N primes}  \\
\hline
\lmnt{p[n]*p[n]>j; p[k++]=j} &
  & \lmnt{p[0] = 2; p[1] = 3; k = 2}
   & \lmnt{A: p[0..k-1] contains the first k primes \&\&
k <= N}  \\
\hline
        &\lmnt{k<N; j = p[k-1]+2; n=0}  &
           & \lmnt{B: A \&\& k<N \&\& relB(p,k,n,j)}                       \\
\hline
\end{tabular}
\caption{\label{fig:primes2}
Next step after Figure~\ref{fig:primes1}:
in column $A$ we have added a transition in column $A$
for the case that {\tt k < N}.
In that case we can start finding the next prime after
{\tt p[k-1]} because we know that there is enough space
in {\tt p} to store it.
{\tt relB(p,k,n,j)} means that
there is no prime between the last prime found and {\tt j}
and that {\tt n<k}, and that {\tt j} is not divided
by any prime in {\tt p[0..n]}.
The missing entry in column {\tt B} leads to a new row and column
labeled {\tt C} in Figure~\ref{fig:primes3}.
}
\end{center}
\end{figure}

In condition {\tt B}
primeness of {\tt j} can be concluded
for sufficiently large {\tt n}.
Initially this is typically not the case,
hence the need for condition {\tt C}:
{\tt B} is true and the square of {\tt p[n]}
is not greater than {\tt j}.
We need to be assured that {\tt n} does not exceed {\tt k-1},
which is a fact of number theory\footnote{
In \cite{vnm11} we pay due attention to such crucial details.
}.

When entering column {\tt C}
we know that {\tt n} is not large enough to conclude
that {\tt j} is prime.
So we need to test for divisibility of {\tt j} by
{\tt p[n+1]}.
Number theory assures
that {\tt p[n+1]} is a prime that has already been found
\cite{vnm11}.
See Figure~\ref{fig:primes3}.

\begin{figure}[htbp]
\begin{center}
\begin{tabular}{|l|l|l|l||l}
\lmnt{C:}& \lmnt{B:} & \lmnt{A:}
     & \lmnt{S: p[0..N-1] exists \&\& N>1} & \\
\hline \hline
& & \lmnt{k >= N} & & \lmnt{H: p[0..N-1] contains the first N primes}  \\
\hline
& \lmnt{p[n]*p[n]>j; p[k++]=j} &
  & \lmnt{p[0] = 2; p[1] = 3; k = 2}
   & \lmnt{A: p[0..k-1] contains the first k primes \&\&
k <= N}  \\
\hline
\lmnt{j\%p[n+1]!=0; n++} & &\lmnt{k<N; j = p[k-1]+2; n=0}  &
           & \lmnt{B: A \&\& k<N \&\& relB(p,k,n,j)}                       \\
\hline
\lmnt{j\%p[n+1]==0; j += 2; n=0}
    & \lmnt{p[n]*p[n] <= j} &
       & & \lmnt{C: B \&\& \\ p[n]*p[n] <= j}\\
\hline
\end{tabular}
\caption{\label{fig:primes3}
Next step after Figure~\ref{fig:primes2}:
Change from Figure~\ref{fig:primes2}:
row and column with label $C$ are added.
There are no incomplete columns,
so this is ready for translation to Java;
see Figure~\ref{fig:javaPrimes}.
}
\end{center}
\end{figure}

This code matrix has no incomplete columns,
so all that remains to be done
is a straightforward translation to Java,
which is possible because all boolean expressions
precede assignments.
See Figure~\ref{fig:javaPrimes}.

\section{Conclusions}

We have introduced Matrix Code,
a hybrid between correctness proof
and program in the form of a matrix with rows and columns
labeled by program assertions.
The entries in the matrix consist of tests or statements
or combinations of both.
One can say that Matrix Code is code-independent.
At the same time it \emph{is} code:
one uses it to program an abstract machine.

The existence of a code-independent notation 
for the algorithm allows us to \emph{separate concerns}.
The concerns are addressed in two stages:
(1) a code matrix from the specification
and (2) compilable code from the code matrix.
Both stages have been demonstrated in this paper.
The second stage is a routine task
because of the special structure of the code matrix.

We have chosen the generation of the prime number table
because the algorithm is around the upper limit of what
most students in a first-year programming course can grasp.
We believe that with matrix code a larger proportion
of the class will be able to approach their programming
assignments not as a trial-and-error process,
but as a goal-directed activity.

\section{Acknowledgments}
We gratefully acknowledge financial support from
the Canadian Natural Sciences and Engineering Research Council
as well as facilities from the University of Victoria.

\begin{figure}[htbp]
\begin{center}
\hrule \vspace{0.1in}
\begin{verbatim}
public static void primesCM(int[] p, int N) {
  final int S=0, A=1, B=2, C=3, H=4;
  int state=S;    // knowledge state
  int j=0, k=0, n=0; // data state
  while (true) {
    switch (state) {
      case S: p[0] = 2; p[1] = 3; k = 2;
        state = A;
      break;
      case A:
        if (k >= N) state = H;
        else {j = p[k-1]+2; n = 0; state = B;}
      break;
      case B: if (p[n]*p[n] > j) {
                p[k++] = j; state = A;
              } else state = C;
      break;
      case C:
        if (j%p[n+1] != 0) {n++; state = B;}
        else {j += 2; n = 0; state = C;}
      break;
      case H: return;
} } }
\end{verbatim}
\end{center}
\caption{\label{fig:javaPrimes}
Translation into a Java function
of the code matrix in Figure~\ref{fig:primes3}.
A gate {\tt b0;S0} in column $X$ and row $R_0$ and
gate {\tt !b0;S1} in column $X$ and row $R_1$
translate to
{\tt case X: if (b0) \{S0; state = R0;\}
else \{S1; state = R1\} break;} in the above code.
We display the literal translation,
before performing the obvious optimizations.
}
\vspace{0.1in}
\hrule
\end{figure}

\end{document}